# INVESTIGATION OF ELECTRIC CONDITIONS IN THE VICINITY OF CARBON NANOTUBES GROWN IN A DC PLASMA SHEATH

*J. Blažek [a], P. Špatenka [b], Ch. Taeschner [c], A. Leonhardt [c]*

[a] University of South Bohemia, Department of Physics, Jeronýmova 10,
CZ-371 15 České Budějovice, Czech Republic

[b] Technical University of Liberec, Department of Materials Science, Hálkova 6,
CZ-461 17 Liberec, Czech Republic

[c] Institut für Festkörper- und Werkstoffforschung Dresden E.V., Helmoltzstrasse 20,
D-01069 Dresden, Germany

**Abstract**
Carbon nanotubes (CNTs – long tubular carbon nanostructures) belong to the best electron field emitting materials. Based on the experimentally determined electron density in the dual hot filament/DC plasma deposition system the electric field close to the substrate and ion energy flux were calculated. The dependences of electric characteristics close to the nanotube tips on various geometric and plasma parameters were numerically investigated. Taking into account the field enhancement in the vicinity of the CNTs tips the force acting on them was determined. The reliability of numeric computations was tested on a model of conductive sphere placed in a homogeneous electric field, for which theoretical predictions concerning electric forces and field emission currents can be derived.

## 1. Introduction

Carbon nanotubes (CNTs) have attracted much research effort since their discovery in 1991 [1]. Their unique structural, chemical, electrical and mechanical properties promise them to be an appropriate material for variety of applications, such as nanoelectronics devices or mechanical reinforcement in composite materials [2,3]. The field emission properties of CNTs offer promising technological applications, i.e. in flat panel displays [4].

Depending on experimental conditions CNTs could grow in one direction (alignment) or in a curly fashion. Only a few attempts have been done to explain the mechanism of the oriented growth, but it is obvious that in a DC plasma sheath the electric field is responsible for the alignment. Yu et al. [5] supposed that the charged particles form bonds along the field lines. Chen et al. [6] expressed a hypothesis that the nanotubes are pulled in the direction of the electric field.

In spite of the hypothesis the CNTs alignment is caused by the electric forces, so far nobody has computed their magnitude. Hence we have developed a code in which both the electric field in the vicinity of CNTs tips and the electric forces acting on them are computed from experimentally measured bulk plasma parameters.

The macroscopic electric field is strongly enhanced at CNT tips and causes cold emission of electrons. Our method of computation of the electric field at the vicinity of CNTs enables us to investigate the dependence of electric current emitted from CNT tips on various geometric parameters such as nanotube density or height.

## 2. Numerical method and results

### 2.1 Computations of electric forces acting on CNT tips

Almost the whole potential bias in the DC plasma is concentrated in a thin layer at the cathode – the sheath. The electric field and electron current satisfy the Poisson and continuity equation. The equation system is completed by an equation connecting the ion velocity with the electric field. Assuming typical working pressure 10 mbar, the collisional ion drift motion [7] has been considered.

Equations mentioned above are due to the CNT geometry 3-dimensional and it is questionable to solve them directly. Fortunately, it can be shown that the electric field is distorted only in the close vicinage of CNTs tips and also the ion current is influenced by the presence of CNTs negligibly. Therefore, it is possible to solve the problem in two steps. In the first step the 3D system of equations is reduced to the 1D (planar) case without CNTs and then solved analytically [7]. In the second step the electric field close to the CNT tips is corrected by the 3D Poisson equation solved only in the vicinity of CNTs, with the ion density taken from the planar case.

From the electric intensity one can easily determine the electric force acting on the nanotube, $F = \varepsilon_0 / 2 \int_S E^2 \cos\theta \, dS$, where $E$ is the intensity just above the nanotube surface. $S$ and $\theta$ is the angle between the normal to the nanotube surface and nanotube longitudinal axis, respectively.

Before computing the forces acting on CNT tips the reliability of implemented method was tested on a sphere placed in the originally homogeneous electric field. The computed value $F = 4.9$ (arbitrary units) in comparison with its exact value 6.3 is fully sufficient for our purposes. For CNTs one can expect even more precise computations, as the geometry is simpler – only one hemisphere is present.

The computations of the field around CNTs were carried out for our experimental conditions, which are described elsewhere [8]. The CNTs were grown in a dual hot filament/DC plasma reactor at the pressure of 10 mbar of hydrogen/methane mixture. The electron number density and electron temperature determined from the probe measurements in the plasma bulk were $2 \times 10^{17}$ m$^{-3}$ and 0.8 eV, respectively. The voltage 650 V between the electrodes was with high accuracy equal to the sheath bias. The radius of the CNTs and the distances between them have been estimated from the SEM images. The "standard" nanotube in our model has the radius of 20 nm and is 70 nm distant from its neighbors. For the sake of simplicity the spherical shape of the catalytic particle on the nanotube tip is supposed. The computations were performed independently for $H^+$ and $CH_3^+$ ions. We present here the numeric values for $CH_3^+$ ions.

The number density of the ions at the cathode was evaluated as $1.8 \times 10^{16}$ m$^{-3}$, their kinetic energy was derived as 4.6 eV. The macroscopic field close to the planar cathode without CNTs was $7.3 \times 10^5$ V/m whereas the maximum field at the top of the nanotube spherical tip was approximately 3 times larger. The resulting value of the electric force acting on the nanotube tip of $1.0 \times 10^{-14}$ N is relatively large and could be really responsible for the alignment. For comparison, this force is equal to the weight of the amorphous carbon cylinder of the same radius and the height of 360 μm and is four order of magnitude higher than the weight of the cobalt droplet ending the nanotube tip.

As the numerical computations show, the force strongly depends on the distances between neighboring nanotubes (Fig.1). When the distance between the nanotubes is smaller than their height, they screen each other and the force is predominately determined by the undisturbed electrical field. With growing distance the field strength increases together with the electrical force. The value of the force is saturated when the distance between cylinders is approximately 5-6 times longer than their height. In such cases the value of the force is the same as for an isolated nanotube.

Fig. 2 shows the relation between the force and the shape of the catalytic droplet on the tip. The droplet was modeled as a rotational ellipsoid with defined ratio of the longitudinal semiaxis to the CNTs radius. The sensitivity of the force to the shape of the CNTs tip depends on the screening effect. The obtained result prompts that for densely distributed nanotubes the force does not notably depend on other forms of the shape, including tiny protrusions on the CNTs tip.

The more detailed explanation of the method and more deeply discussion and interpretation of numeric results will be published elsewhere [9].

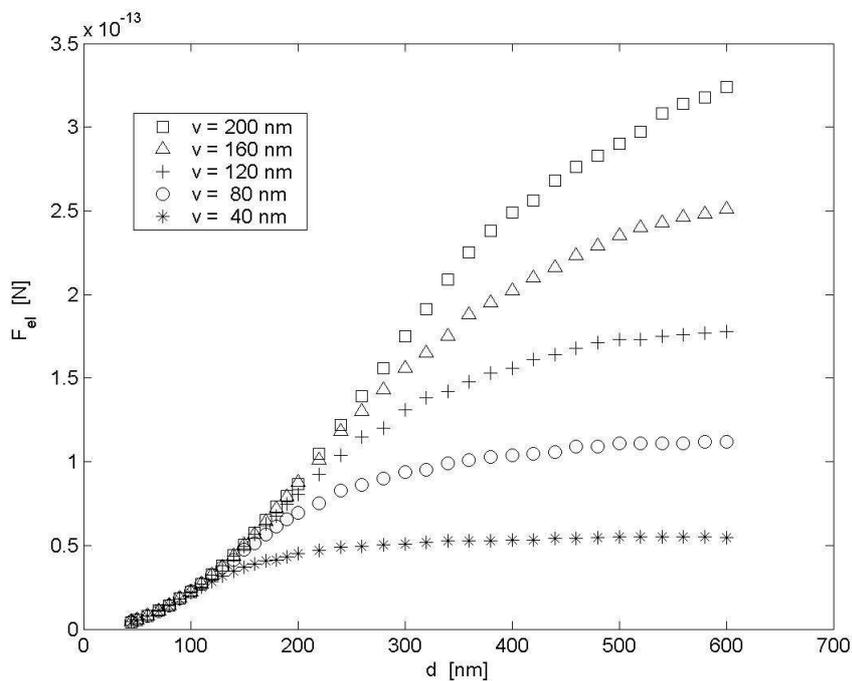

**Fig. 1** Electric force $F_{el}$ versus distance $d$ between two neighboring nanotubes (radius 20 nm, spherical tip).

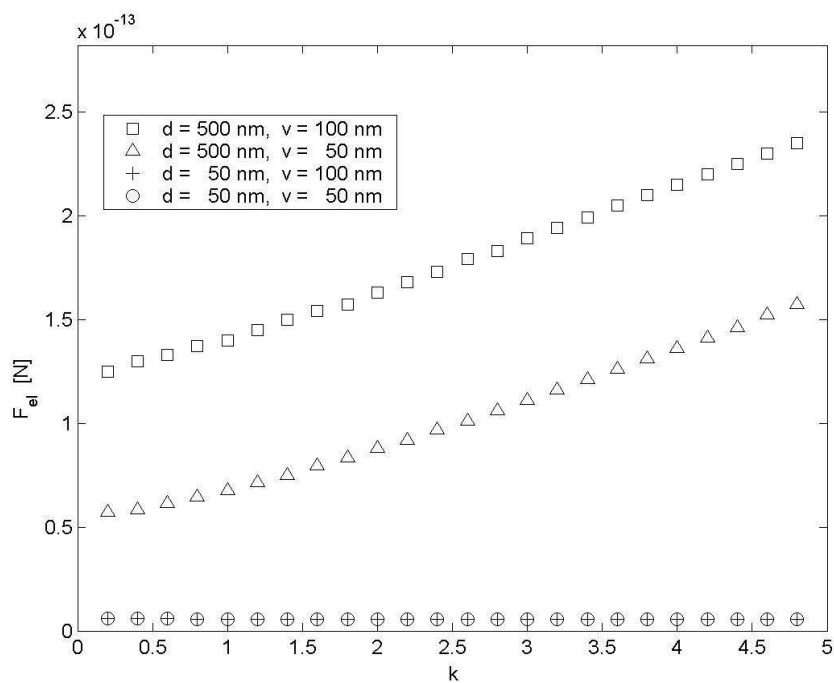

**Fig.2** Electric force $F_{el}$ s a function of the ratio $k = c/r$, where $c$ is the longitudinal semiaxis of the rotational ellipsoidal droplet on the CNTs tip and $r = 20$ nm is its radius; the curves are computed for two different distances $d$ between the nanotubes and two different heights $v$.

## 2.2 Computations of field electron emission from carbon nanotubes

The field electron emission is in the first approximation described by Fowler-Nordheim theory [10]. The electric field above the emitting surface enables the electrons tunneling outward. The serious problem arises in connection with the local field $E$ close to the emitter tip, as it cannot be measured directly. It is expressed as $E = \gamma E_0$, where $E_0$ is the average macroscopic field, $E_0 = U/d$, where $U$ is the voltage between negative substrate and positive electrode and $d$ is their distance. Coefficient $\gamma$ is the field enhancement factor. To compute this coefficient we have applied a numeric method developed for computation of electric forces. To verify the numerical reliability of our computations, we again tested them on a conductive sphere placed in a homogeneous field. The field enhancement factor $\gamma$ of a sphere can be theoretically evaluated, $\gamma = 3\cos\theta$, where $\theta$ is the angle between the normal of the surface and the homogeneous field. Therefore, the total FE current emitted from the sphere can be expressed analytically. The computations are in agreement with theoretical predictions within the relative error of about 20%.

In computations of the field enhancement factor the screening effect arising from neighboring nanotubes has to be taken into account. The screening effect is inversely proportional to the distances of the neighboring nanotubes. For high-density film the emission from one nanotube is low whereas the density of CNTs emitters is high. On the contrary, for low-density film the emission from one tip is high and the density of emitters is low. Hence, there exists an optimum distance of CNTs, for which the electron current density reaches its maximum. This effect was experimentally established and theoretically explained in [11]. We investigated this effect numerically. Fig.2 depicts the macroscopic current density (current per unit of area) versus the distance. For large distances the current density drops as $1/d^2$.

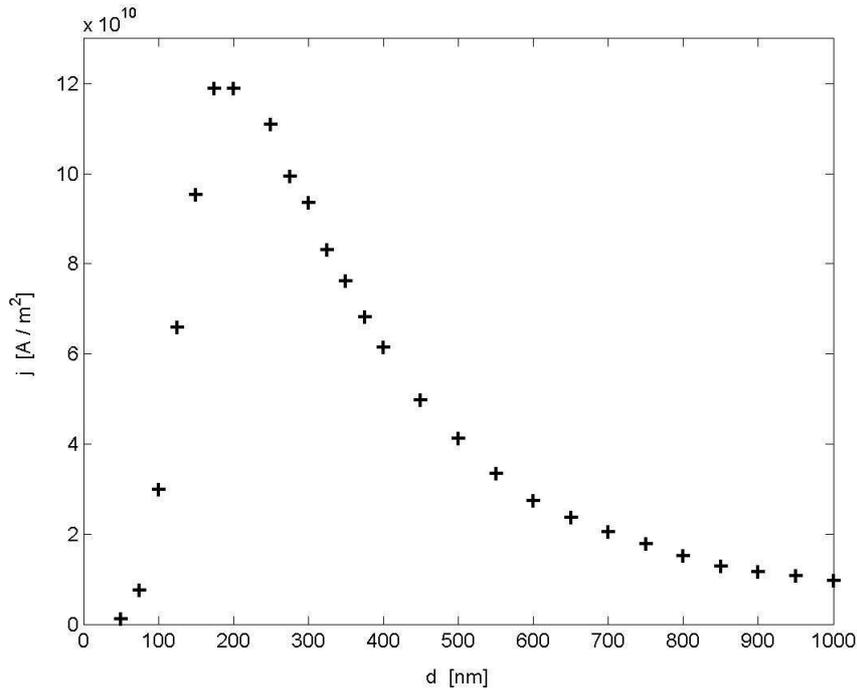

**Fig. 3** Current density vs. distance; macroscopic intensity $E_0 = 5\cdot10^9$ V/m.

The computational model also shows that – contrary to the force – the FE current is strongly sensitive to the shape of the CNTs tips (Fig. 4). We again modeled the nanotube tip as a hemi-ellipsoidal surface characterized by the ratio $k = c/R$, where $c$ and $R$ are the axial semiaxis and radius of the nanotube, respectively. For lower intensities the FE current is more sensitive to the shape of the emitting area than for larger ones. This phenomenon can be probably explained by the existence of an effective emission area. For stronger field the emission area is approximately constant and equal to the whole hemi-ellipsoidal surface. For lower field the emission current density is significant only at some vicinity of the apex, where the local intensity exceeds some threshold value. Hence, the increase of the parameter *k* is accompanied by increase of both the local field and emitting area and the emission current changes more rapidly.

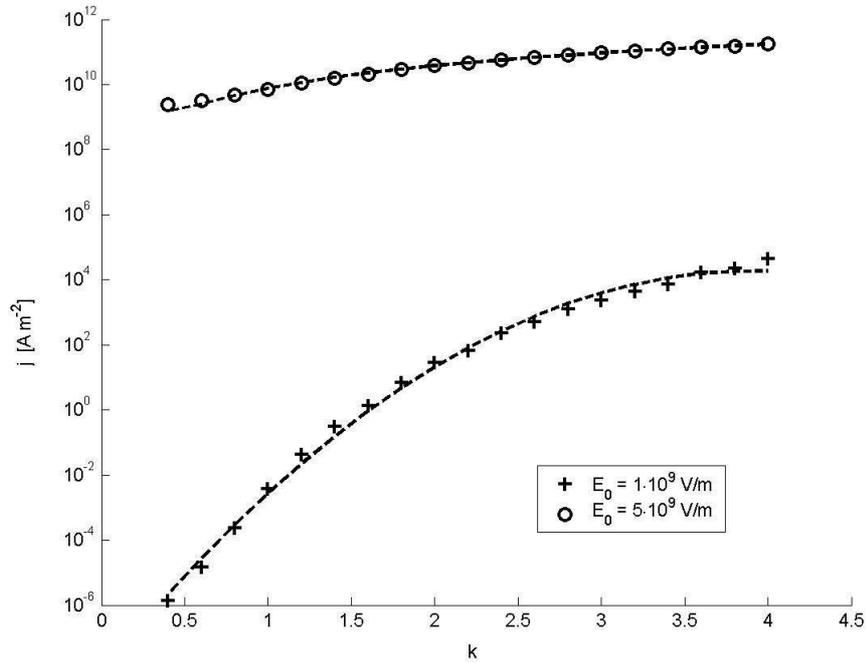

**Fig. 4** Dependence of the current emitted from one nanotube with the ellipsoidal tip on the ratio $k \equiv c/R$, where *c* is the vertical semiaxis and *R* is the nanotube radius.

## 3. Conclusion

We have developed a complex model enabling evaluation of various plasma and electric characteristics near the nanotube tips grown in a DC collisional plasma sheath. It has been shown that electric forces acting on CNT tips are serious candidates for nanotube alignment. The computations of the electric field emission from the nanotube tips have confirmed that there exists some optimum density of nanotubes for which the emitted current reaches its maximum.

**Acknowledgement:**
This work was supported by the projects MSM 124100004 and OC 527.60.